\documentclass[aps,prl,twocolumn,showpacs,byrevtex]{revtex4}

\usepackage{times,mathptm}
\usepackage[dvips]{graphicx,color}
\usepackage{titlesec}
\newcommand{\bqa}{\begin{eqnarray*}}
\newcommand{\eqa}{\end{eqnarray*}}

\begin{document}

\title{Pressure dependence of the superconducting transition temperature of compressed LaH$_{10}$}
\author{Chongze Wang, Seho Yi, and Jun-Hyung Cho$^{*}$}
\affiliation{
Department of Physics, Research Institute for Natural Science, and HYU-HPSTAR-CIS High Pressure Research Center, Hanyang
University, 222 Wangsimni-ro, Seongdong-Ku, Seoul 04763, Republic of Korea}

\date{\today}

\begin{abstract}
Two recent experiments [M. Somayazulu $et$ $al$., Phys. Rev. Lett. ${\bf 122}$, 027001 (2019) and A. P. Drozdov $et$ $al$., Nature ${\bf 569}$, 528 (2019)] reported the discovery of superconductivity in the fcc phase of LaH$_{10}$ at a critical temperature $T_{\rm c}$ between 250${\sim}$260 K under a pressure of about 170 GPa. However, the dependence of $T_{\rm c}$ on pressure showed different patterns: i.e., the former experiment observed a continuous increase of $T_{\rm c}$ up to ${\sim}$275 K on further increase of pressure to 202 GPa, while the latter one observed an abrupt decrease of $T_{\rm c}$ with increasing pressure. Here, based on first-principles calculations, we reveal that for the fcc-LaH$_{10}$ phase, softening of the low-frequency optical phonon modes of H atoms dramatically occurs as pressure decreases, giving rise to a significant increase of the electron-phonon coupling (EPC) constant. Meanwhile, the electronic band structure near the Fermi energy is insensitive to change with respect to pressure. These results indicate that the pressure-dependent phonon softening is unlikely associated with Fermi-surface nesting, but driven by effective screening with the electronic states near the Fermi energy. It is thus demonstrated that the strong variation of EPC with respect to pressure plays a dominant role in the decrease of $T_{\rm c}$ with increasing pressure, supporting the measurements of Drozdov $et$ $al$.
\end{abstract}

\maketitle

In 1968, Neil Ashcroft proposed that a metallic solid-hydrogen could exhibit superconductivity (SC) at high temperatures~\cite{MetalicH_Ashc}. This metallic hydrogen can be referred to as a conventional Bardeen-Cooper-Schrieffer (BCS) superconductor, where SC is driven by a condensate of electron pairs, so-called Cooper pairs, due to electron-phonon interactions~\cite{BCS}. Despite such a theoretical proposal of high-temperature SC, the experimental realization of metallic hydrogen has been very challenging, because it requires too high pressures over ${\sim}$400 GPa~\cite{MetalicH1,MetalicH2,MetalicH3}. In order to achieve the metallization of hydrogen at relatively lower pressures attainable in diamond anvil cells~\cite{diamondanvil,diamondanvil2}, an alternative route has been taken by using hydride materials in which hydrogen atoms can be ``chemically precompressed" to reduce the distances between neighboring H atoms~\cite{chemically-precompressed1,chemically-precompressed2,chemically-precompressed3}. This route with hydrides has recently been demonstrated to be promising for the achievement of room-temperature SC~\cite{room-temperature-sc1,room-temperature-sc2,room-temperature-sc3} that is one of the most challenging subjects in modern physics. Nearly simultaneously, two experimental groups synthesized a lanthanum hydride LaH$_{10}$ with a clathrate-like structure [see Fig. 1(a)] at megabar pressures and measured a superconducting transition temperature ($T_{\rm c}$) between 250${\sim}$260 K at a pressure of ${\sim}$170 GPa~\cite{ExpLaH10-1,ExpLaH10-2}. Obviously, this record of $T_{\rm c}$ is the highest temperature so far among experimentally available superconducting materials, thereby opening a new era of high-temperature SC~\cite{ExpH3S,Hydride1,Hydride2,Hydride3,Hydride4}.

\begin{figure}[htb]
\centering{ \includegraphics[width=8.5cm]{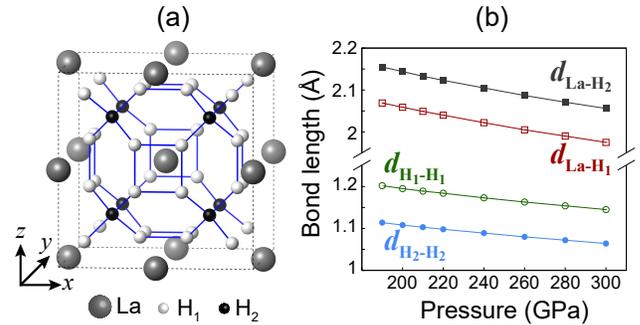} }
\caption{(Color online) (a) Optimized structure of the fcc phase of compressed LaH$_{10}$. The H-rich clathrate structure LaH$_{10}$ is composed of the H$_{32}$ cage surrounding a La atom. The two different types of H atoms, i.e., H$_1$ and H$_2$, are drawn with bright and dark circles, respectively. The positive $x$, $y$, and $z$ axes point along the [001], [010], and [001] directions, respectively. (b) Calculated H$_1$-H$_1$, H$_1$-H$_2$, La-H$_1$, and La-H$_2$ bond lengths of the fcc LaH$_{10}$ structure as a function of pressure.}
\end{figure}

Although the two experiments~\cite{ExpLaH10-1,ExpLaH10-2} agreed well with high values of $T_{\rm c}$ in the fcc phase of compressed LaH$_{10}$, the dependence of $T_{\rm c}$ on pressure is conflicting with each other. The measurements of Somayazulu $et$ $al$.~\cite{ExpLaH10-1} showed a continuous increase of $T_{\rm c}$ up to ${\sim}$275 K at 202 GPa, while those of Drozdov $et$ $al$.~\cite{ExpLaH10-2} displayed a dome-like shape of the $T_{\rm c}$ vs. pressure values, which represents an abrupt decrease of $T_{\rm c}$ with increasing pressure over ${\sim}$170 GPa. The latter experimental measurements are similar to the theoretical prediction of a previous density-functional-theory (DFT) calculation~\cite{room-temperature-sc1}, where $T_{\rm c}$ of the fcc LaH$_{10}$ phase decreases with increasing pressure in the range between 210 and 300 GPa. Meanwhile, it is expected that, as pressure increases, the lattice constants and the H$-$H bond lengths of fcc LaH$_{10}$ decrease. The resulting stronger covalent bonding of H atoms should increase the frequencies of H-derived phonons, which in turn contribute to enhance $T_{\rm c}$ in terms of the standard weak-coupling BCS expression~\cite{MetalicH_Ashc,BCS}. Therefore, the two different experimental observations~\cite{ExpLaH10-1,ExpLaH10-2} on the pressure dependence of  $T_{\rm c}$ of fcc LaH$_{10}$ raise an open question of whether $T_{\rm c}$ increases or decreases with increasing pressure, together with its microscopic underlying mechanism.

In this Rapid Communication, using first-principles DFT calculations, we investigate the bonding, electronic, and phononic properties of fcc LaH$_{10}$ as a function of pressure to identify the pressure dependence of $T_{\rm c}$ with its microscopic mechanism. It is found that the Fermi surface and the van Hove singularity (vHs) near the Fermi energy $E_{\rm F}$, formed by the holelike and electronlike bands with a strong hybridization of the La 4$f$ and H $s$ orbitals~\cite{liangliang-prb2019}, is insensitive to change with respect to pressure. By contrast, the low-frequency optical phonon modes of H atoms shift dramatically toward lower frequencies with decreasing pressure, leading to an overlap with the acoustic phonon modes of La atoms. Such softening of the low-frequency optical phonon modes at lower pressures is thus unlikely associated with Fermi-surface nesting, but is due to an enhanced electron-phonon coupling (EPC) with the electronic states at the vHs. These unique features of phonons and electronic band structure with respect to pressure give rise to a significant increase of the EPC constant at lower pressures, resulting in a nearly linear decrease of $T_{\rm c}$ with increasing pressure. Therefore, the present results support the measurements of Drozdov $et$ $al$.~\cite{ExpLaH10-2} on the $T_{\rm c}$ vs. pressure relation of fcc LaH$_{10}$.

\begin{figure}[ht]
\includegraphics[width=8.5cm]{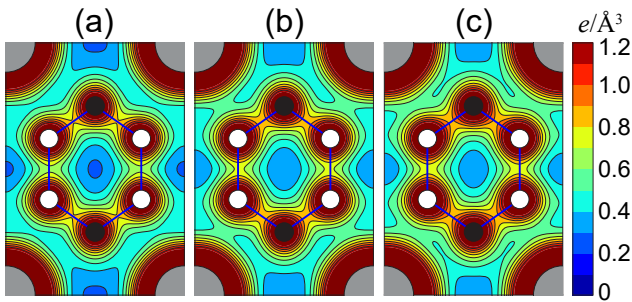}
\caption{(Color online) Calculated total charge densities of fcc LaH$_{10}$ at (a) 220 GPa, (b) 260 GPa, and (c) 300 GPa. The charge density contour maps are drawn in the (1$\bar{1}$0) plane with a contour spacing of 0.1 $e$/{\AA}$^3$.}
\end{figure}

We first optimize the structure of the fcc LaH$_{10}$ phase as a function of pressure using the DFT calculations~\cite{methods}. Based on previous experimental~\cite{ExpLaH10-1,ExpLaH10-2} and theoretical~\cite{room-temperature-sc1,room-temperature-sc2,room-temperature-sc3} studies, we consider the fcc LaH$_{10}$ structure having cages of 32 H atoms surrounding a La atom [see Fig. 1(a)]~\cite{room-temperature-sc1,room-temperature-sc2}. It is noted that there are two types of H atoms: i.e., H$_1$ atoms forming the squares and H$_2$ atoms forming the hexagons. The optimized fcc LaH$_{10}$ structures as a function of pressure show that the lattice constants decrease monotonously with increasing pressure (see Fig. S1 of the Supplemental Material~\cite{supple}). Consequently, as shown in Fig. 1(b), the H$_1$-H$_1$, H$_1$-H$_2$, La-H$_1$, and La-H$_2$ bond lengths (denoted as $d_{\rm H_1-H_1}$, $d_{\rm H_1-H_2}$, $d_{\rm La-H_1}$, and $d_{\rm La-H_2}$, respectively) also decrease monotonously with increasing pressure. At 300 GPa, we obtain $d_{\rm H_1-H_1}$ = 1.145 {\AA} and $d_{\rm H_1-H_2}$ = 1.064 {\AA}, in good agreement with those ($d_{\rm H_1-H_1}$ = 1.152 and $d_{\rm H_1-H_2}$ = 1.071 {\AA}) of a previous DFT calculation~\cite{room-temperature-sc1}. Figures 2(a), 2(b), and 2(c) show the calculated total charge densities of fcc LaH$_{10}$ at 220. 260, and 300 GPa, respectively. It is seen that, as pressure increases, the charge densities at the midpoints of the H$_{1}$-H$_1$ and H$_{1}$-H$_2$ bonds increase, indicating that their covalent bonding characters increase. Such enhanced H-H covalent-bond strengths with increasing pressure should contribute to increase the frequencies of H-derived phonons, as discussed below. It is noticeable that there is also a covalent character between La atoms and H$_{32}$ cages~\cite{liangliang-prb2019}: i.e., the electrical charges of La and H$_1$ atoms are connected with each other at 260 and 300 GPa. As shown in Figs. 2(a), 2(b), and 2(c), the charge densities between La and H$_1$ atoms decrease with decreasing pressure. This reduced covalent strength between La atoms and H$_{32}$ cages is well reflected by the increase of the bond lengths $d_{\rm La-H_1}$ and $d_{\rm La-H_2}$ with decreasing pressure [see Fig. 1(b)], which will later be shown to be related with softening of the low-frequency optical phonon modes of H atoms. It is thus likely that the increases of $d_{\rm La-H_1}$ and $d_{\rm La-H_2}$ lead to induce the dynamical instability of the fcc LaH$_{10}$ phase, as observed by experiments~\cite{ExpLaH10-1,ExpLaH10-2}.

\begin{figure}[htb]
\includegraphics[width=8.5cm]{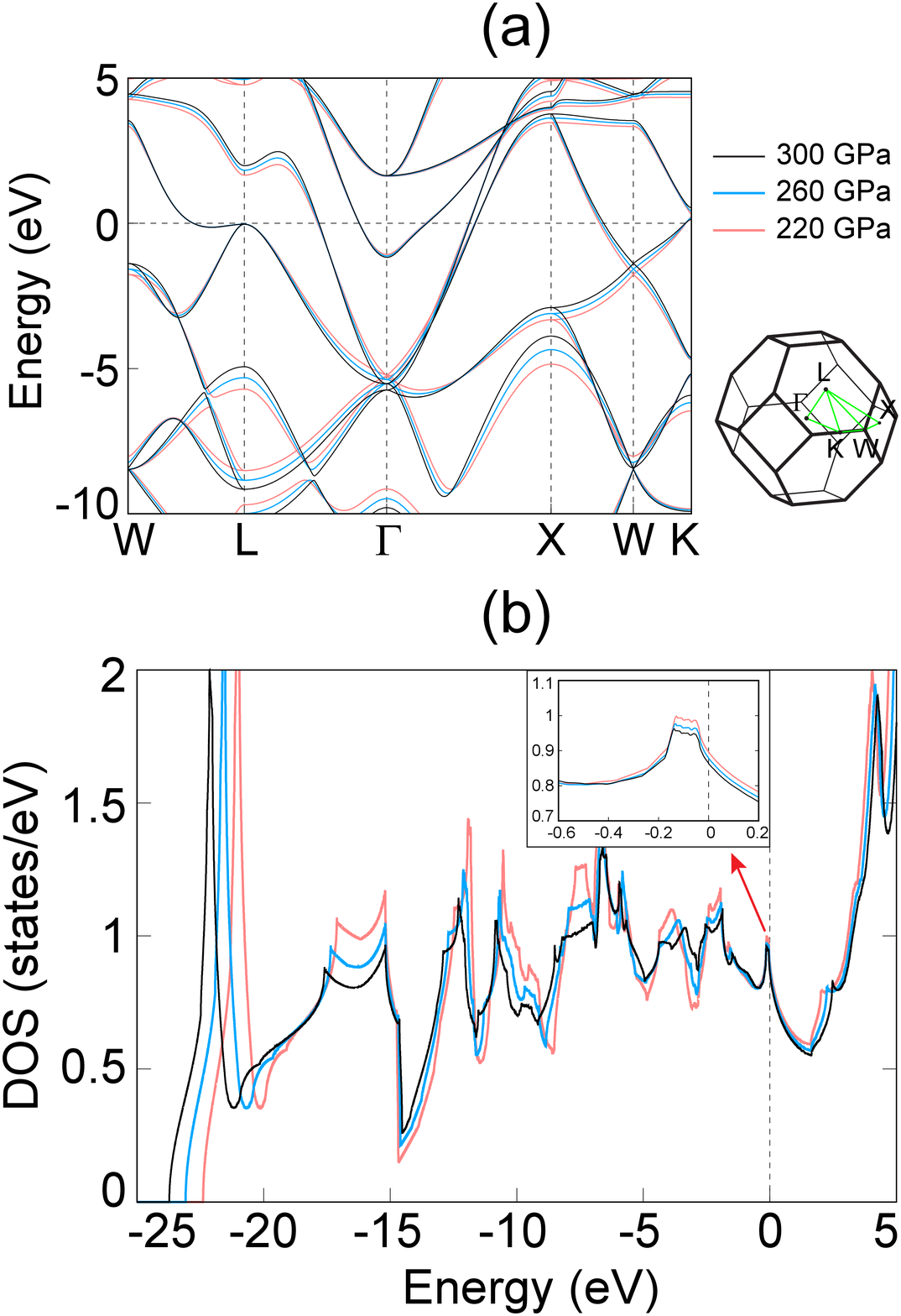}
\caption{(Color online) Comparison of (a) the electronic band structures and (b) DOS of fcc LaH$_{10}$, calculated at 220, 260, and 300 GPa. The energy zero represents $E_{\rm F}$. In (a), the Brillouin zone of the fcc primitive cell is also drawn. The inset of (b) shows a closeup of the DOS around the vHs near $E_{\rm F}$. }
\end{figure}

Figure 3(a) shows the comparison of the electronic band structures of fcc LaH$_{10}$, calculated at 220, 260, and 300 GPa. Here, we consider three different pressures 220, 260, and 300 GPa, because the experimentally observed fcc LaH$_{10}$ phase becomes stable above a critical pressure between 210 and 220 GPa, as discussed below. We find that the dispersions of the bands around $E_{\rm F}$ change very little with respect to pressure, indicating a nearly invariance of the Fermi surface. Therefore, as shown in Fig. 3(b), the pressure dependence of the density of states (DOS) around $E_{\rm F}$ is minor compared to those at the energy regions away from $E_{\rm F}$. It is noted that the double-shaped vHs, produced by the presence of the holelike and electronlike bands around the high-symmetry $L$ points~\cite{liangliang-prb2019}, exhibits a slight decrease of the DOS at $E_{\rm F}$ with increasing pressure [see the inset of Fig. 3(b)]. According to our previous DFT calculations~\cite{liangliang-prb2019}, the existence of such vHs near $E_{\rm F}$ is of importance for the room-temperature SC observed~\cite{ExpLaH10-1,ExpLaH10-2} in fcc LaH$_{10}$. Here, the electronic states composing the vHs have a strong hybridization of the La 4$f$ and H$_1$ $s$ orbitals~\cite{liangliang-prb2019}. These hybridized electronic states near $E_{\rm F}$ not only contribute to produce the covalent character between La and H$_1$ atoms (as discussed above) but also could effectively screen the low-frequency optical phonon modes of H atoms, the frequencies of which are significantly lowered as pressure decreases (as discussed below). In contrast to the bands near $E_{\rm F}$, the valence and conduction bands away from $E_{\rm F}$ shift downwards and upwards with increasing pressure, respectively [see Figs. 3(a) and 3(b)]. Such shifts of the valence and conduction bands are likely caused by the band widening due to increased electron hopping at higher pressures. As shown in Fig. 3(b), the calculated DOS shows that the bottom of the valence bands is lowered from $-$22.4 eV at 220 GPa to $-$23.7 eV at 300 GPa.

\begin{figure*}[ht]
\includegraphics[width=17cm]{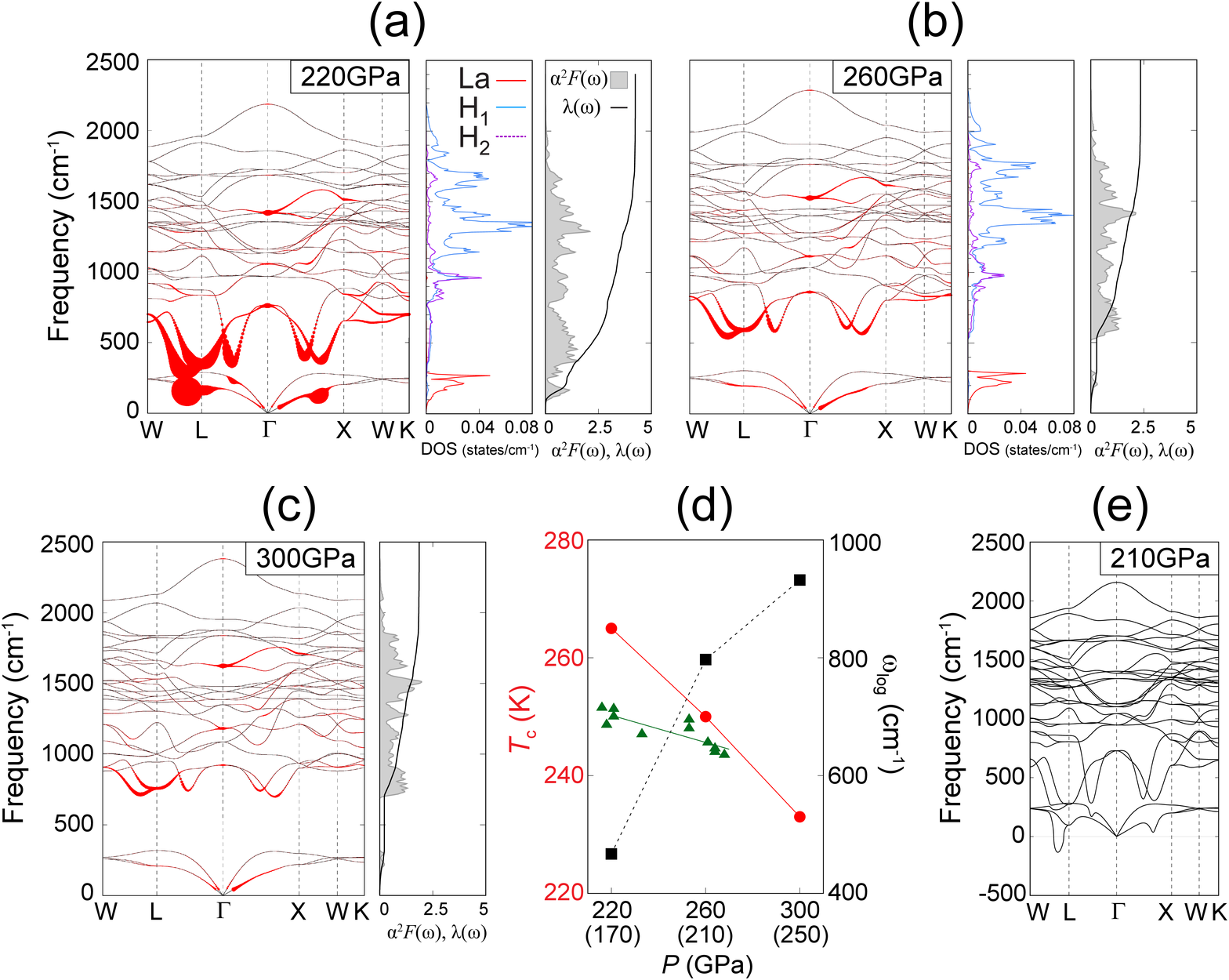}
\caption{(Color online) Calculated phonon spectrum, phonon DOS projected onto La, H$_1$, and H$_2$ atoms, Eliashberg function ${\alpha}^{2}F({\omega})$, and integrated EPC constant ${\lambda}({\omega})$ of fcc LaH$_{10}$ at (a) 220 GPa, (b) 260 GPa, and (c) 300 GPa. Here, the size of circles on the phonon dispersion is proportional to the EPC strength. (d) Calculated $T_{\rm c}$ (displayed with circles) and ${\omega}_{\rm log}$ (squares) of fcc LaH$_{10}$ as a function of pressure. For comparison, the experimentally observed $T_{\rm c}$ of Drozdov $et$ $al$.~\cite{ExpLaH10-2} between ${\sim}$170 and ${\sim}$220 GPa are given in (d), where the scale of experimental pressure is given in parentheses. (e) Calculated phonon spectrum of fcc LaH$_{10}$ at 210 GPa. }
\end{figure*}

Next, we study the phonon spectrum of fcc LaH$_{10}$ as a function of pressure using the density functional perturbation theory implemented in QUANTUM
ESPRESSO~\cite{QE}. Figures 4(a), 4(b), and 4(c) dispaly the calculated phonon dispersions at 220, 260, and 300 GPa, respectively, together with the projected DOS onto La, H$_1$, and H$_2$ atoms. We find that at 300 GPa, the acoustic phonon modes of La atoms with frequencies lower than ${\sim}$315 cm$^{-1}$ are well separated from the optical phonon modes of H atoms with frequencies higher than ${\sim}$700 cm$^{-1}$ [see Fig. 4(c)]. However, as pressure decreases, the H-derived optical modes shift downwards in the overall frequency range. Specifically, at 220 GPa, the low-frequency optical modes around the $L$ and $X$ points significantly shift to lower frequencies, thereby overlapping with the acoustic modes [see Fig. 4(a)]. Since the Fermi surface changes very little with respect to pressure, such softening of the low-frequency optical modes at lower pressures is likely due to their effective screening with the electronic states near $E_{\rm F}$, rather than Fermi-surface nesting. It is noted that, as pressure decreases, the increase in the bond lengths of $d_{\rm La-H_1}$ and $d_{\rm La-H_2}$ would induce the effective screening of the low-frequency optical modes with the strongly hybridized La 4$f$ and H$_1$ $s$ electronic states around the high-symmetry $L$ points at $E_{\rm F}$. Consequently, the EPC strength in the low-frequency optical modes around the $L$ point is much enhanced with decreasing pressure, as illustrated by the circles in Fig. 4(a). Here, the larger the size of circle, the stronger is the EPC.

Figures 4(a), 4(b), and 4(c) also include the Eliashberg function ${\alpha}^{2}F({\omega})$ and the integrated EPC constant ${\lambda}({\omega})$ as a function of phonon frequency. We find that the contributions to ${\alpha}^{2}F({\omega})$ and ${\lambda}({\omega})$ arise from all three phonon modes including the La-derived acoustic, the H$_1$-derived optical, and the H$_2$-derived optical modes. Here, since the projected phonon DOS onto La atoms is well separated from those of H$_1$ and H$_2$ atoms, we can estimate that at 260 (300) GPa, the acoustic phonon modes of La atoms contribute to ${\sim}$12 (12)\% of the total EPC constant ${\lambda}$ = ${\lambda}$(${\infty}$), while the optical phonon modes of H$_1$ and H$_2$ contribute to ${\sim}$61 and ${\sim}$27\% (${\sim}$65 and ${\sim}$23\%) of ${\lambda}$, respectively. Therefore, the contributions of the H$_1$-derived optical modes to ${\lambda}$ are larger than those of the H$_2$-derived optical and the La-derived acoustic modes. As shown in Figs. 4(a), 4(b), and 4(c), ${\lambda}$ decreases with increasing pressure as 4.24, 2.35, and 1.86 at 220, 260, and 300 GPa, respectively. Meanwhile, the logarithmic average of phonon frequencies, ${\omega}_{\rm log}$, increases with increasing pressure as 467, 797, and 932 cm$^{-1}$ at 220, 260, and 300 GPa, respectively [see Fig. 4(d)]. These opposite variations of ${\lambda}$ and ${\omega}_{\rm log}$ with respect to pressure are combined to slightly influence the change of $T_{\rm c}$ for the two pressure ranges between 220-260 GPa and 260-300 GPa. Here, ${\lambda}$ (${\omega}_{\rm log}$) contributes to decrease (increase) $T_{\rm c}$ with increasing pressure~\cite{ADnote}. By numerically solving the Eliashberg equations~\cite{Eliash} with the typical Coulomb pseudopotential parameter of ${\mu}^*$ = 0.13~\cite{room-temperature-sc1,room-temperature-sc2}, we find that $T_{\rm c}$ decreases with increasing pressure as 265, 250, and 233 K at 220, 260, and 300 GPa, respectively [see Figs. 4(d) and S2 of the Supplemental Material~\cite{supple}]. Thus, we can say that ${\lambda}$ plays a more dominant role in determining the pressure dependence of $T_{\rm c}$, rather than ${\omega}_{\rm log}$. It is noticeable that at 210 GPa, the negative phonon frequencies appear along the $L-W$ line [see Fig. 4(e)], indicating that the fcc LaH$_{10}$ phase becomes unstable. Thus, our results of the $T_{\rm c}$ vs. pressure relation in fcc LaH$_{10}$ show that $T_{\rm c}$ decreases by ${\sim}$16 K for every 40-GPa increase above the critical pressure $P_{\rm c}$ between 210 and 220 GPa [see Fig. 4(d)]. This linear slope of $T_{\rm c}$ as a function of pressure agrees well with a previous DFT result~\cite{room-temperature-sc1}, but it is larger than the experimental measurements of Drozdov $et$ $al$.~\cite{ExpLaH10-2} where $T_{\rm c}$ decreases with a slope of 6${\pm}$1 K between 170 and 210 K. It is noticeable that the experimental critical pressure $P_{\rm c}$ ${\approx}$ 170 GPa~\cite{ExpLaH10-2,LaH10-expt-angew} is much lower than the present and previous~\cite{room-temperature-sc1} theoretical ones between 210 and 220 GPa. This difference of $P_{\rm c}$ between experiment and theory was explained by the anharmonic effects on phonons~\cite{LaH10-expt-angew,LaH10-anharmonic-prb}. Thus, in order to better reproduce the observed~\cite{ExpLaH10-1,ExpLaH10-2} pressure dependence of $T_{\rm c}$ in fcc LaH10, further theoretical investigations with including anharmonic or nonadiabatic effects~\cite{H3S-anharmonic-1,H3S-anharmonic-2,H3S-nonadiabatic} are demanded in the future.

In summary, our first-principles calculations have demonstrated that $T_{\rm c}$ of the compressed fcc LaH$_{10}$ phase decreases with increasing pressure. It is revealed that, as pressure decreases, the low-frequency optical phonon modes of H atoms are dramatically softened due to their effective screening with the electronic states near $E_{\rm F}$, which are characterized by a strong hybridization of the La 4$f$ and H$_1$ $s$ orbitals. Interestingly, the electronic band structure around $E_{\rm F}$ changes very little with respect to pressure. Therefore, such unique features of phonons and electronic band structure with respect to pressure result in a large increase of the EPC constant with decreasing pressure. Our findings provided a microscopic understanding of why $T_{\rm c}$ of fcc LaH$_{10}$ decreases with increasing pressure, supporting the recent experimental measurements of Drozdov $et$ $al$.~\cite{ExpLaH10-2}. The present explanation for the pressure dependence of $T_{\rm c}$ of fcc LaH$_{10}$ based on the strong variation of EPC with respect to pressure is rather generic and hence, it could be more broadly applicable to other high-$T_{\rm c}$ hydrides with structural instability. Indeed, compressed H$_3$S having $T_{\rm c}$ = 203 K at ${\sim}$150 GPa~\cite{ExpH3S} was also observed to exhibit a decrease of $T_{\rm c}$ with increasing pressure.

\vspace{0.4cm}

\noindent {\bf Acknowledgement.}
This work was supported by the National Research Foundation of Korea (NRF) grant funded by the Korean Government (Grants No. 2019R1A2C1002975, No. 2016K1A4A3914691, and No. 2015M3D1A1070609). The calculations were performed by the KISTI Supercomputing Center through the Strategic Support Program (Program No. KSC-2018-CRE-0063) for the supercomputing application research.  \\

\noindent $^{*}$ Corresponding author: chojh@hanyang.ac.kr

\newpage
\onecolumngrid
\titleformat*{\section}{\LARGE\bfseries}

\renewcommand{\thefigure}{S\arabic{figure}}
\setcounter{figure}{0}

\vspace{1.2cm}

\section{Supplemental Material for "Pressure dependence of the superconducting transition temperature of compressed LaH$_{10}$"}
\vspace{1.2cm}
\begin{flushleft}
{\bf 1. Comparison of the lattice constants and electronic bands of fcc LaH10, obtained using a PAW+PBE pseudopotential in VASP and an ONCV+PBE pseudopotential in QE.}
\begin{figure}[ht]
\includegraphics[width=16cm]{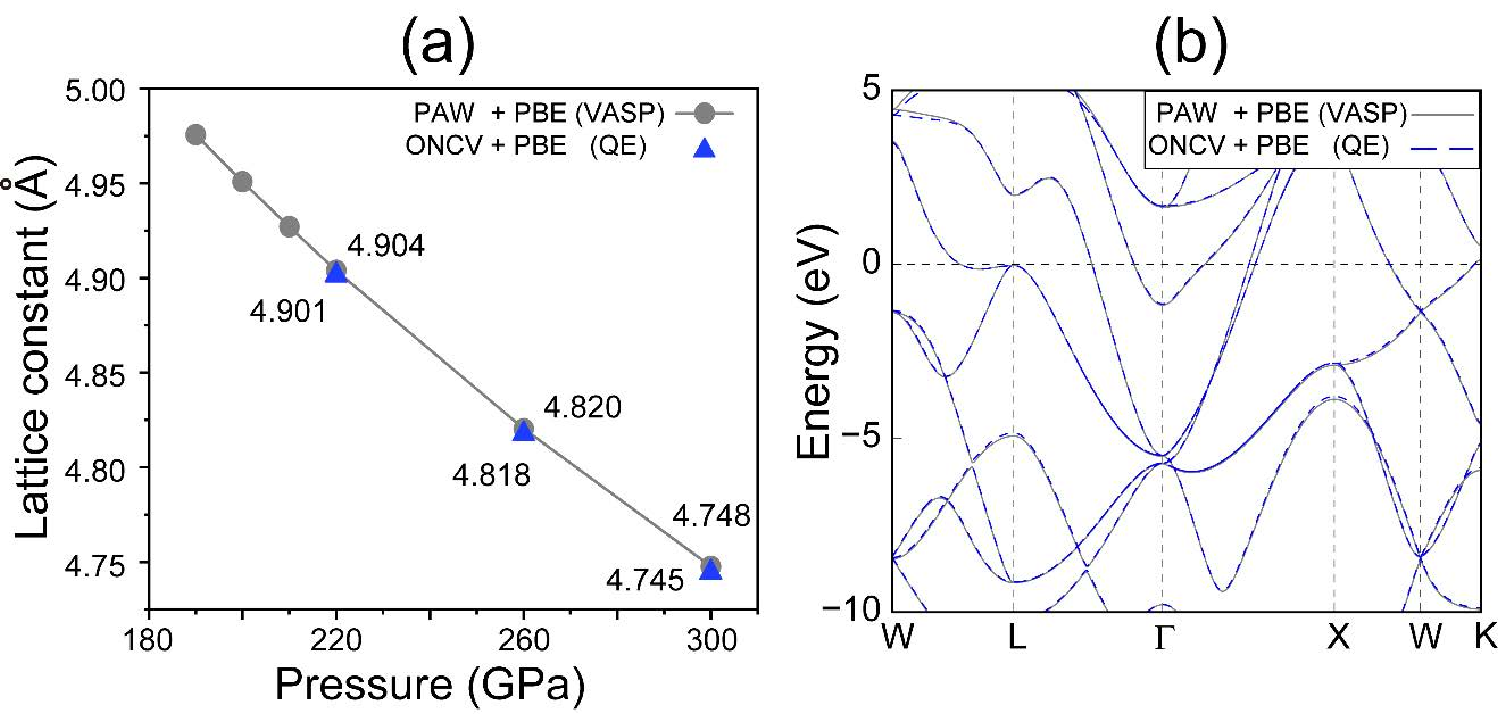}
\caption{ (a) Optimized lattice constants $a$ = $b$ = $c$ of fcc LaH$_{10}$ using a PAW+PBE pseudopotential in VASP (grey circles) and an ONCV+PBE pseudopotential in QE (blue triangles). The calculated values (in {\AA}) at 220, 260, and 300 GPa are also given in (a). (b) Calculated electronic band structures of fcc LaH$_{10}$ at 300 GPa, obtained using the two pseudoptentials. }
\end{figure}

\vspace{1.2cm}

{\bf 2. Superconducting energy gap of fcc LaH$_{10}$ at different pressures.}
\begin{figure}[ht]
\includegraphics[width=8cm]{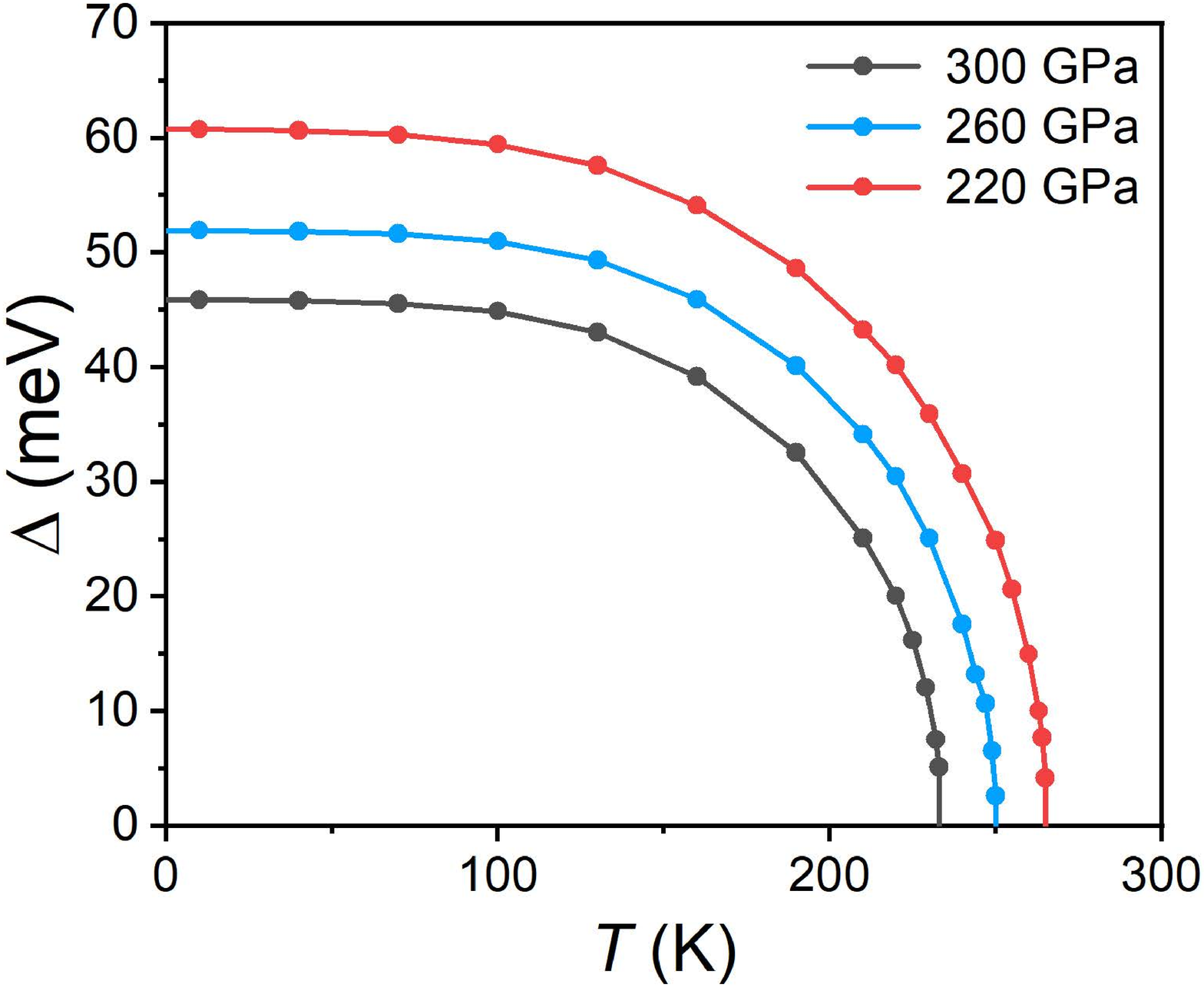}
\caption{ Superconducting energy gap ${\Delta}$ as a function of temperature $T$, obtained at 220, 260, and 300 GPa. Here, we used the typical Coulomb
pseudopotential parameter of ${\mu}^*$ = 0.13.}
\end{figure}
\vspace{0.4cm}

{\bf 3. Convergence test of ${\lambda}$ with respect to $k$-point and $q$-point meshes.}
\begin{figure}[ht]
\includegraphics[width=8cm]{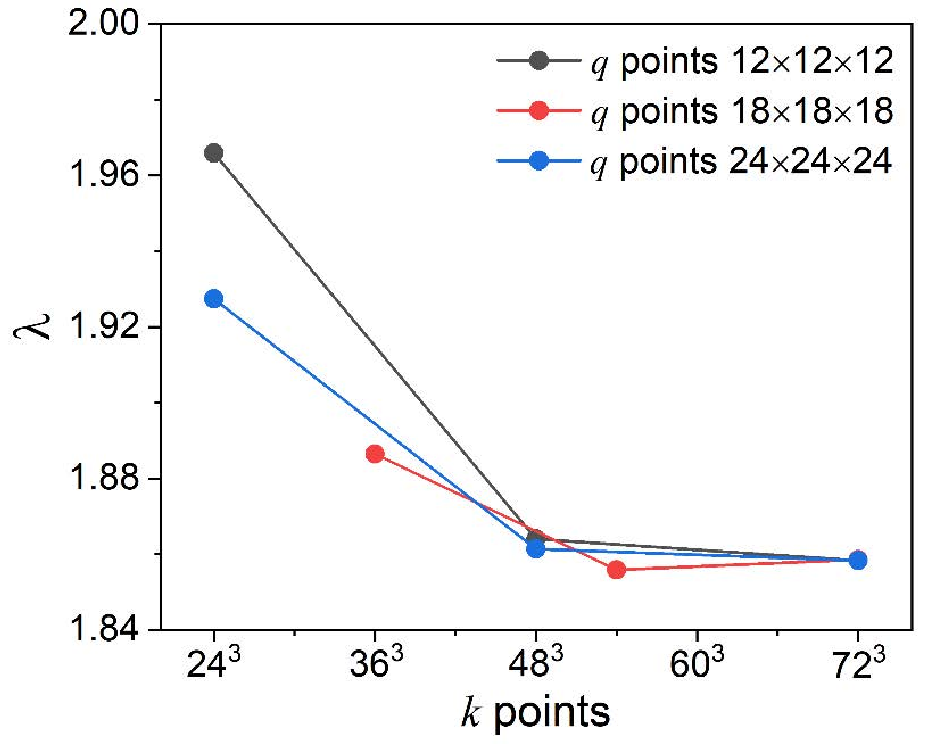}
\caption{Calculated ${\lambda}$ of fcc LaH$_{10}$ as functions of $k$ points and $q $ points. Here, the pressure is taken at 300 GPa.}
\end{figure}
\end{flushleft}
\end{document}